\documentclass[traditabstract,oldversion]{aa}

\usepackage{amsmath}
\usepackage{graphicx}
\usepackage{multirow}
\usepackage{txfonts}
\usepackage[authoryear]{natbib}
\usepackage{lscape}
\usepackage{rotating}
\usepackage{ifthen}
\usepackage{hyperref}

\bibpunct{(}{)}{;}{a}{}{,}

\begin{document}

\title{An improved map of the Galactic Faraday sky}

\author{N.~Oppermann\thanks{\email{niels@mpa-garching.mpg.de}} \inst{\ref{inst:MPA}}
	\and H.~Junklewitz \inst{\ref{inst:MPA}}
	\and G.~Robbers \inst{\ref{inst:MPA}}
	\and M.R.~Bell \inst{\ref{inst:MPA}}
	\and T.A.~En\ss{}lin \inst{\ref{inst:MPA}}
	\and A.~Bonafede \inst{\ref{inst:Jacobs}}
	\and R.~Braun \inst{\ref{inst:ATNF}}
	\and J.C.~Brown \inst{\ref{inst:calgary}}
	\and T.E.~Clarke \inst{\ref{inst:NRL}}
	\and I.J.~Feain \inst{\ref{inst:ATNF}}
	\and B.M.~Gaensler \inst{\ref{inst:sydney}}
	\and A.~Hammond \inst{\ref{inst:sydney}}
	\and L.~Harvey-Smith \inst{\ref{inst:ATNF}}
	\and G.~Heald \inst{\ref{inst:ASTRON}}
	\and M.~Johnston-Hollitt \inst{\ref{inst:VUW}}
	\and U.~Klein \inst{\ref{inst:AIFA}}
	\and P.P.~Kronberg \inst{\ref{inst:toronto},\ref{inst:LANL}}
	\and S.A.~Mao \inst{\ref{inst:ATNF},\ref{inst:CFA}}
	\and N.M.~McClure-Griffiths \inst{\ref{inst:ATNF}}
	\and S.P.~O'Sullivan \inst{\ref{inst:ATNF}}
	\and L.~Pratley \inst{\ref{inst:VUW}}
	\and T.~Robishaw \inst{\ref{inst:DRAO}}
	\and S.~Roy \inst{\ref{inst:NCRA}}
	\and D.H.F.M.~Schnitzeler \inst{\ref{inst:ATNF},\ref{inst:MPIFR}}
	\and C.~Sotomayor-Beltran \inst{\ref{inst:AIRUB}}
	\and J.~Stevens \inst{\ref{inst:ATNF}}
	\and J.M.~Stil \inst{\ref{inst:calgary}}
	\and C.~Sunstrum \inst{\ref{inst:calgary}}
	\and A.~Tanna \inst{\ref{inst:UNSW}}
	\and A.R.~Taylor \inst{\ref{inst:calgary}}
	\and C.L.~Van~Eck \inst{\ref{inst:calgary}}}

\institute{Max Planck Institute for Astrophysics, Karl-Schwarzschild-Str.~1, 85741 Garching, Germany \label{inst:MPA}
	\and Jacobs University Bremen, Campus Ring 1, 28759 Bremen, Germany \label{inst:Jacobs}
	\and Australia Telescope National Facility, CSIRO Astronomy \& Space Science, PO Box 76, Epping, NSW 1710, Australia \label{inst:ATNF}
	\and Department of Physics \& Astronomy, University of Calgary, AB T2N 1N4, Canada \label{inst:calgary}
	\and Naval Research Laboratory, 4555 Overlook Ave.~SW, Washington, DC 20375, USA \label{inst:NRL}
	\and Sydney Institute for Astronomy, School of Physics, The University of Sydney, NSW 2006, Australia \label{inst:sydney}
	\and ASTRON, Postbus 2, 7990 AA Dwingeloo, The Netherlands \label{inst:ASTRON}
	\and School of Chemical \& Physical Sciences, Victoria University of Wellington, PO Box 600, Wellington 6140, New Zealand \label{inst:VUW}
	\and Argelander-Institut f\"ur Astronomie, Universit\"at Bonn, Auf dem H\"ugel 71, 53121 Bonn, Germany \label{inst:AIFA}
	\and Department of Physics, University of Toronto, 60 St George Street, Toronto ON M5S 1A7, Canada \label{inst:toronto}
	\and Los Alamos National Laboratory, M.S.~B283, Los Alamos, NM 87545, USA \label{inst:LANL}
	\and Harvard-Smithsonian Center for Astrophysics, Cambridge, MA 02138, USA \label{inst:CFA}
	\and Dominion Radio Astrophysical Observatory, HIA-NRC, 717 White Lake Road, P.O. Box 248, Penticton, BC, V2A 6J9, Canada \label{inst:DRAO}
	\and National Centre for Radio Astrophysics, TIFR, Pune University Campus, Post Bag 3, Ganeshkhind, Pune 411007, India \label{inst:NCRA}
    \and Max Planck Institut f\"ur Radioastronomie, Auf dem H\"ugel 69, 53121 Bonn, Germany \label{inst:MPIFR}
	\and Astronomisches Institut der Ruhr-Universit\"at Bochum, Universit\"atsstra{\ss}e 150, 44780 Bochum, Germany \label{inst:AIRUB}
	\and School of Physics, University of New South Wales, Sydney NSW 2052, Australia \label{inst:UNSW}}

\date{Received DD MMM. YYYY / Accepted DD MMM. YYYY}

\abstract{We aim to summarize the current state of knowledge regarding Galactic Faraday rotation in an all-sky map of the Galactic Faraday depth. For this we have assembled the most extensive catalog of Faraday rotation data of compact extragalactic polarized radio sources to date. In the map making procedure we use a recently developed algorithm that reconstructs the map and the power spectrum of a statistically isotropic and homogeneous field while taking into account uncertainties in the noise statistics. This procedure is able to identify some rotation angles that are offset by an integer multiple of $\pi$. The resulting map can be seen as an improved version of earlier such maps and is made publicly available, along with a map of its uncertainty. For the angular power spectrum we find a power law behavior $C_\ell\propto\ell^{-2.17}$ for a Faraday sky where an overall variance profile as a function of Galactic latitude has been removed, in agreement with earlier work. We show that this is in accordance with a 3D Fourier power spectrum $P(k)\propto k^{-2.17}$ of the underlying field $n_\mathrm{e}B_r$ under simplifying geometrical and statistical assumptions.
}

\keywords{Galaxies: magnetic fields - Galaxy: structure - ISM: magnetic fields - Radio continuum: ISM - Methods: data analysis - Techniques: polarimetric}

\titlerunning{The Galactic Faraday sky}
\authorrunning{N.~Oppermann et al.}
\maketitle

\section{Introduction}

Magnetic fields are ubiquitous in the interstellar medium. They are likely to play a major dynamical role in the evolution of galaxies. It is by comparing theoretical predictions and simulations to observations of galactic magnetic fields that their generation and dynamical role can be understood \citep[see e.g.][and references therein]{beck-2011}. It is natural to look first and foremost at our own galaxy, the Milky Way, and try to study its magnetic field. However, its observation is complicated by a number of effects. The magnetic field is a three-dimensional vector field that varies on multiple scales throughout the Galaxy. Thus, a large number of measurements of the field would be needed to determine even its large-scale properties. Furthermore, virtually any observation suffers from a projection effect as local effects add up along the line of sight. And finally the magnetic field cannot be measured directly, so that related observables have to be used. These observables, however, are not only sensitive to the magnetic field itself but also to other quantities which are not necessarily better understood, introducing ambiguities when inferring properties of the magnetic field. The intensity of synchrotron radiation is sensitive to the strength of the magnetic field component orthogonal to the line of sight, however it is modulated by the density of cosmic ray electrons \citep[e.g.][]{ginzburg-1965}. The direction of this magnetic field component can be studied via the polarization direction of synchrotron radiation and thermal dust emission \citep[e.g.][]{gardner-1966, lazarian-2003}. A magnetic field component along the line of sight, on the other hand, gives rise to the effect of Faraday rotation \citep[e.g.][]{nicholson-1983, gardner-1966, burn-1966}. The strength of this effect is influenced not only by the magnetic field but also by the density of thermal electrons. Furthermore, when observing this effect for extragalactic sources, it contains contributions non only from the Galaxy, but rather from every position along the line of sight to the source with a non-vanishing magnetic field and thermal electron density.

In order to find an unambiguous terminology capturing these subtleties, we introduce the concept of Faraday depth, which depends on position and is independent of any astrophysical source. The Faraday depth corresponding to a position at a distance $r_0$ from an observer is given by a line of sight integral,
\begin{equation}
	\phi(r_0)=\frac{e^3}{2\pi m_e^2c^4}\int_{r_0}^0\mathrm{d}r~n_\mathrm{e}(r)B_r(r),
\end{equation}
over the thermal electron density $n_\mathrm{e}$ and the line of sight component of the magnetic field $B_r$. Here, $e$ and $m_e$ are the electron charge and mass and $c$ is the speed of light. The Galactic Faraday depth is therefore exactly this integral, where the lower boundary is the outer edge of the Milky Way. It is this integral that contains the information on the Galactic magnetic field.

The observational consequence of Faraday rotation on a single linearly polarized source is a rotation of its plane of polarization about an angle that is proportional to the square of the wavelength. The proportionality constant is equal to the source's Faraday depth, i.e the above integral expression, where the lower boundary is now the source's position. Often, the assumption that the observed polarized radiation stems from a single source is made implicitly and a linear fit to the position angle of the plane of polarization as a function of the squared wavelength is made. We refer to the slope of such a $\lambda^2$-fit as rotation measure (RM). In the case of a single source this is the same as the source's Faraday depth. However, the polarized radiation will in general be emitted over a range of physical distances and also over a range of Faraday depths, and the position angle will no longer vary linearly with $\lambda^2$. This emission spectrum in Faraday space can be recovered using the technique of RM synthesis \citep{burn-1966, brentjens-2005}. In this work we create a map of the Galactic Faraday depth using both data that are based on RM synthesis and data that are based on linear $\lambda^2$-fits. Neither measures the Galactic Faraday depth exclusively and we use the term Faraday rotation data when referring to data values without specifying whether they are rotation measures or the result of a synthesis study.

A review of early work on the inference of features of the regular component of the Galactic magnetic field from RM measurements is included in the work of \citet{frick-2001}. Some of the studies are done by \citet{morris-1964, gardner-1969, vallee-1973, ruzmaikin-1977, ruzmaikin-1978, simard-normandin-1979, andreasian-1980, andreasian-1982, inoue-1981, sofue-1983, vallee-1983a, agafonov-1988, clegg-1992, han-1994, han-1997}, as well as \citet{rand-1989, rand-1994} who use RM data of pulsars, and \citet{seymour-1966, seymour-1984} who uses spherical harmonics to obtain an all-sky RM map. Some of the more recent studies aiming to constrain the Galactic magnetic field using rotation measures of extragalactic radio sources include the ones by \citet{brown-2001, mao-2010, kronberg-2011, pshirkov-2011}, as well as \citet{brown-2003b, brown-2007, nota-2010, vaneck-2011}, who supplement extragalactic RMs with pulsar rotation measures. \citet{weisberg-2004, vallee-2005, vallee-2008, han-2006, men-2008} rely entirely on pulsar rotation measures for estimating the Galactic magnetic field, while \citet{sun-2008, jansson-2009, jaffe-2010} use rotation measures of extragalactic sources in combination with synchrotron polarization and intensity data.

Recent attempts to create an all-sky map of Faraday rotation measure were made by \citet{frick-2001,johnston-hollitt-2004a,dineen-2005,xu-2006}. However, due to the limited number of data points available at the time, their reconstructions are limited to the largest-scale features. A rather sophisticated attempt is made by \citet{short-2007}, who use Monte Carlo Markov Chain methods and account for uncertainty in the noise covariance while avoiding the direct involvement of covariance matrices. Realistic attempts to create all-sky maps including smaller-scale features have been possible only since \citet{taylor-2009} published the NRAO VLA Sky Survey (NVSS) \citep{condon-1998} rotation measure catalog that contains data on sources distributed roughly equally over the sky at declinations larger than $-40^\circ$. One such attempt is made in the same publication where the data are simply smoothed to cover the celestial sphere in regions where data are taken. Another attempt has been made by \citet{oppermann-2011a}, using a more sophisticated signal reconstruction algorithm which takes into account spatial correlations without oversmoothing any maxima or minima.

The NVSS rotation measure catalog is, however, suboptimal in two respects. It lacks data in a large region in the southern sky below the declination of $-40^\circ$ due to the position of the observing telescope (VLA) and its rotation measure values were deduced using only two nearby frequency channels (see Table~\ref{tab:datasets}). This increases the risk of introducing offsets of integer multiples of $\pi$ in the rotation angle, as discussed by \citet{sunstrum-2010}, and makes it impossible to detect any deviations from a proportionality to $\lambda^2$ in the polarization angle. Thus, sources with a non-trivial Faraday spectrum could not be identified and were assigned a possibly misleading RM value.

In this work we aim to create a map of the Galactic Faraday depth that summarizes the current state of knowledge. To this end we combine the NVSS rotation measure catalog of \citet{taylor-2009} with several other catalogs of Faraday rotation data of polarized extragalactic radio sources, increasing the spatial coverage and further constraining the signal also in regions where several data sets overlap. We improve on the map of \citet{oppermann-2011a} by using this more extensive data set and by using an extended version of the reconstruction algorithm which takes into account uncertainties in the noise covariance, presented by \citet{oppermann-2011b}. The resulting all-sky map of the Galactic Faraday depth will be useful in many respects. On the one hand, all-sky information can help in bringing forth global features of the underlying physics, such as the Galactic magnetic field or the electron distribution. On the other hand, an all-sky map can also be useful when studying local or extragalactic features. It could, for example, serve as a look-up table for Galactic contributions to the Faraday depth when studying extragalactic objects.

The remainder of this paper is organized as follows: In Sect.~\ref{sec:algorithm} we briefly review the main features of the extended critical filter algorithm that we use in our map making procedure and discuss how it is applied to the situation at hand. The data sets entering the reconstruction are listed in Sect.~\ref{sec:datasets} and the results are presented in Sect.~\ref{sec:results}. In the results section, we also include a brief discussion of the reconstructed angular power spectrum. We summarize our findings in Sect.~\ref{sec:conclusions}.

\section{Reconstruction algorithm}
\label{sec:algorithm}

In order to reconstruct the Galactic Faraday depth from the point source measurements, we use the \textit{extended critical filter} formalism that was presented by \citet{oppermann-2011b}. This filter is based on the \textit{critical filter} that was used for the reconstruction by \citet{oppermann-2011a} and derived by \citet{ensslin_frommert-2011} and \citet{ensslin_weig-2010} within the framework of \textit{information field theory} developed by \citet{ensslin-2009}.

\subsection{Signal model}

The signal model we use is the same as the one used by \citet{oppermann-2011a}. We review the essentials briefly.

In the inference formalism we employ, it is assumed that a linear relationship, subject to additive noise, exists between the observed data $d$ and the signal field $s$ that we try to reconstruct, i.e.
\begin{equation}
	d=Rs+n.
\end{equation}
Here, the response operator $R$ describes the linear dependence of the data onto the signal. Formally, the signal could be a continuous field, e.g.\ some field like the Galactic Faraday depth on the celestial sphere. In practice, however, the best we can hope for is to reconstruct a discretized version of such a field, i.e.\ a pixelized sky-map. In this case, one can think of the signal field $s$ on the sphere as a vector of dimension $N_\mathrm{pixels}$, each component of which corresponds to one pixel, and the whole set of data points $d$ as another vector of dimension $N_\mathrm{data}$. The response operator then becomes a matrix of dimension $N_\mathrm{data}\times N_\mathrm{pixels}$ and $n$ is another vector of dimension $N_\mathrm{data}$ that contains the noise contributions to each data point. Next, we specify the definitions of the signal field and the response matrix for our specific application.

The critical filter algorithm, as well as the extended critical filter, is intended to reconstruct statistically isotropic and homogeneous random signal fields. We briefly recapture the meaning of this.

It is assumed in the derivation of the filter formulas \citep[see][for details]{oppermann-2011b}, that the signal field that describes nature is one realization of infinitely many possible ones. Further, it is assumed that some of these possibilities are a priori more likely to be realized in nature than others, i.e.\ a prior probability distribution function on the space of all possible signal realizations is defined. We assume this probability distribution to ba a multivariate Gaussian with an autocorrelation function $S(\hat{n},\hat{n}')$. Here, $\hat{n}$ and $\hat{n}'$ denote two positions on the celestial sphere. Now assuming statistical homogeneity and isotropy means assuming that $S(\hat{n},\hat{n}')$ depends only on the angle between the two positions $\hat{n}$ and $\hat{n}'$. This means that the correlation of the value of the signal field at one position with another one at a certain distance depends only on this distance, not on the position on the sphere (homogeneity) and not on the direction of their separation (isotropy). Note, however, that we are making this assumption only for the prior probability distribution, i.e.\ the inherent probability for signal realizations. The data can (and do) break this symmetry, making the posterior probability distribution, i.e.\ the probability for a signal realization given the measured data, anisotropic. Furthermore, any single realization of a signal with isotropic statistics can appear arbitrarily anisotropic. Extremely anisotropic realizations will, however, be a priori more unlikely than others.

For this reason we divide out the most obvious largest scale anisotropy introduced by the presence of the Galactic disk. We do this by defining our signal as
\begin{equation}
	s(l,b)=\frac{\phi(l,b)}{p(b)},
\end{equation}
i.e.\ the dimensionless ratio of the Galactic Faraday depth $\phi$ and a variance profile $p$ that is a function of Galactic latitude only. We use this simplistic ansatz for the Galactic variance profile in order to account for the largest scale anisotropies without using any specific Galactic model in the analysis.

\begin{figure}
	\input{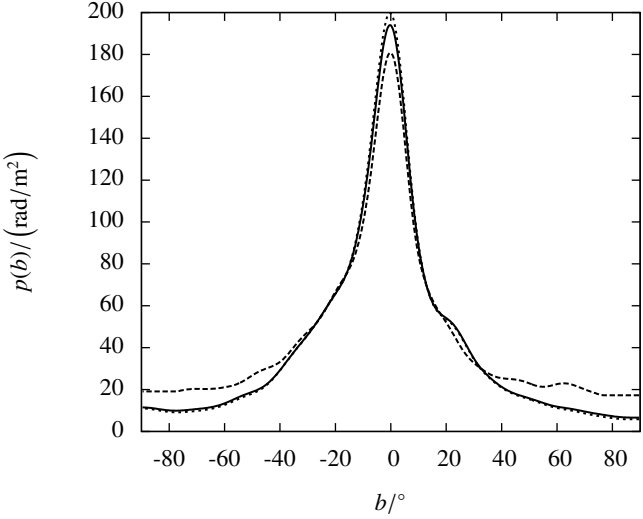}
	\caption{\label{fig:profile}The root mean square Galactic profile that is used in the definition of the signal field and is part of the response matrix, as a function of Galactic latitude. The dashed curve represents the initial profile function $\tilde{p}$ and the solid curve the one used in the final reconstruction, $p$. The dotted curve shows the profile as calculated from the final results.}
\end{figure}

The profile function is calculated in a multi-step procedure. In the first step, we sort the data points into bins of Galactic latitude and calculate the root mean square value for the Faraday rotation data of each bin, disregarding any information on Galactic longitude of the data points. We then smooth these values with a kernel with $10^\circ$ FWHM\footnote{\citet{oppermann-2011a} experiment with different smoothing lengths and find that a factor two difference does not matter for the end result. We chose $10^\circ$ by visual inspection of the smoothness of the resulting profile.} to form an initial profile function $\tilde{p}$. In the second step, we reconstruct the signal field, resulting in a map $\tilde{m}$ and the corresponding $1\sigma$ uncertainty map $\hat{\tilde{D}}^{1/2}$. We use these to calculate the corresponding posterior mean of the squared Faraday depth according to
\begin{equation}
	\left<\phi^2\right>_{\mathcal{P}(s|d)}=\tilde{p}^2\tilde{m}^2+\tilde{p}^2\hat{\tilde{D}}.
\end{equation}
The posterior mean is the ensemble average over all possible signal configurations weighted with their posterior probability distribution $\mathcal{P}(s|d)$, i.e.\ their probability given the measured data, and is denoted by $\left<\cdot\right>_{\mathcal{P}(s|d)}$. From this expected map of the squared Faraday depth, we then calculate a new variance profile $p$, now using the pixel values of the map instead of the data points. A few data points were added before repeating this final step yet another time. The final reconstruction is then conducted with the resulting profile function. Both the initial variance profile and the one used in the final reconstruction are shown in Fig.~\ref{fig:profile}. The drop-off toward the Galactic poles of the first-guess profile function is less pronounced since the relatively high noise component of the Faraday rotation data in these regions enters in the root mean square that is calculated from the data points. The variance profile as calculated from the final results is also shown in Fig.~\ref{fig:profile}.

Having introduced the Galactic variance profile, we can now specify the response operator. In our application, the response matrix $R$ needs to contain both the multiplication of the signal field with this profile function and the probing of the resulting Faraday depth in the directions of the point sources. It is a matrix of dimension $N_\mathrm{data}\times N_\mathrm{pixels}$. Each row corresponds to one data point and each column to one pixel of the sky map. Here, the row corresponding to the $i$-th data point contains a non-zero entry only in the column corresponding to the pixel in which the $i$-th observed extragalactic source lies, modeling the probing of the Faraday depth in the observed directions. This entry is the value of the Galactic variance profile $p$ at the latitude of the pixel, effectively rescaling the local signal field value into a Faraday depth.

Furthermore, we assume Gaussian priors both for the signal and for the noise with covariance matrices $S$ and $N$, respectively. Since our signal field is assumed to be statistically homogeneous and isotropic, its covariance matrix $S$ is completely determined by its angular power spectrum\footnote{The angular power spectrum is defined by $C_\ell=\left<s_{\ell m}s_{\ell m}^*\right>_{\mathcal{P}(s)}$, where $s_{\ell m}$ denotes the signal's spherical harmonic component of a certain azimuthal quantum number $\ell$ and an arbitrary magnetic quantum number $m$, the asterisk denotes complex conjugation, and the angular brackets denote an ensemble average weighted with the prior probability distribution.} $(C_\ell)_\ell$, $\ell=0,1,\dots,\ell_\mathrm{max}$. The minimum length scale $\ell_\mathrm{max}$ is determined by the finite resolution of the discretization. Assuming uncorrelated noise for all data points, the noise covariance $N$ becomes diagonal. The diagonal entries are given by the variance calculated from the error bars given in the data catalogs, modified to account for the expected average extragalactic contribution,
\begin{equation}
	\sigma^2=\sigma_\mathrm{(measurement)}^2+\sigma_\mathrm{(extragalactic)}^2.
	\label{eq:sigmacorr}
\end{equation}
We include a multiplicative correction factor $\eta$ that will be determined during the reconstruction, making the diagonal entry of $N$ corresponding to the $i$-th data point
\begin{equation}
	N_{ii}=\eta_i\sigma_i^2.
\end{equation}
As the extragalactic contribution, we use the value $\sigma_\mathrm{(extragalactic)}=6.6~\mathrm{rad}/\mathrm{m}^{2}$, motivated by the study of \citet{schnitzeler-2010}.

Reasons for a deviation of $\eta$ from unity could be a general under-estimation of the measurement error, as was discussed for the NVSS catalog by \citet{stil-2011}, a misestimation of the extragalactic contribution, a multi-component Faraday depth spectrum, but also the presence of an offset of an integer multiple of $\pi$ in the rotation angle.

\subsection{The extended critical filter}

The extended critical filter \citep[see][]{oppermann-2011b} is a method to simultaneously reconstruct the signal, its covariance, given here by its angular power spectrum $(C_\ell)_\ell$, and the noise covariance, given here by the correction factors $(\eta_i)_i$. To this end, inverse Gamma distributions are assumed as priors for the parameters of the covariances, i.e.
\begin{equation}
	\mathcal{P}(C_\ell)=\frac{1}{q_\ell\Gamma(\alpha_\ell-1)}\left(\frac{C_\ell}{q_\ell}\right)^{-\alpha_\ell}\exp\left(-\frac{q_\ell}{C_\ell}\right)
\end{equation}
and
\begin{equation}
	\mathcal{P}(\eta_i)=\frac{1}{r_i\Gamma(\beta_i-1)}\left(\frac{\eta_i}{r_i}\right)^{-\beta_i}\exp\left(-\frac{r_i}{\eta_i}\right),
	\label{eq:etaprior}
\end{equation}
and all these parameters are assumed to be independent. We choose $\alpha_\ell=1$ for the parameter describing the slope of the power law and $q_\ell=0$ for the parameter giving the location of the exponential low-amplitude cutoff, turning the prior for each $C_\ell$ into Jeffreys prior which is flat on a logarithmic scale, enforcing the fact that we have no a priori information on the power spectrum. For the prior of the correction factors we choose the parameter $\beta_i=2$, since we already have information on the expected noise covariance from the data catalogs. We adapt the value of $r_i$ such that the a priori expectation value of $\log\eta$ becomes $0$, thereby conforming with the catalogs.

With these values, the actual filtering process consists of iterating the three equations\footnote{This is the first order version of the extended critical filter. See \citet{oppermann-2011b} for details.}
\begin{equation}
	\label{eq:WF}
	m=DR^\dagger N^{-1}d,
\end{equation}
\begin{equation}
	\label{eq:Cl}
	C_\ell=\frac{1}{2\ell+1}\mathrm{tr}\left(\left(mm^\dagger+D\right)S_\ell^{-1}\right),
\end{equation}
and
\begin{equation}
	\label{eq:eta}
	\eta_i=\frac{1}{2\beta_i-1}\left[2r_i+\frac{1}{\sigma_i^2}\left(\left(d-Rm\right)_i^2+\left(RDR^\dagger\right)_{ii}\right)\right]
\end{equation}
until convergence is reached. Here, $m$ is the reconstructed signal map, the $\dagger$-symbol denotes a transposed quantity, and $D=\left(S^{-1}+R^\dagger N^{-1}R\right)^{-1}$ is the so-called information propagator \citep{ensslin-2009}. The matrix $S_\ell^{-1}$ projects a signal vector onto the $\ell$-th length-scale by keeping only the degrees of freedom represented by spherical harmonics components with the appropriate azimuthal quantum number. Although we have chosen $\beta_i=2$ for our reconstruction, we leave the parameter unspecified in these equations, since we later compare our results to those obtained with $\beta\neq2$ (see Sect.~\ref{sec:noise-rec}).

The three equations can be qualitatively explained. Eq.~\eqref{eq:WF} links the reconstructed map to the data. It consists of a response over noise weighting of the data and an application of the information propagator to the result. The information propagator combines knowledge about the observational procedure encoded in the response matrix $R$ and the noise covariance matrix $N$ with information on the signal's correlation structure contained in the signal covariance matrix $S$. It is used in Eq.~\eqref{eq:WF} to reconstruct the map at a given location by weighting the contributions of all data points using this information. The information propagator is also (approximatively) the covariance matrix of the posterior probability distribution. Therefore, it can be used to obtain a measure for the uncertainty of the map estimate. The 1$\sigma$ uncertainty of the map estimate in the $j$-th pixel is given by $\hat{D}^{1/2}_j=D_{jj}^{1/2}$. Eq.~\eqref{eq:Cl} estimates the angular power spectrum from two contributions. The first term in the trace gives the power contained within a reconstructed map, while the second term compensates for the power lost in the filtering procedure generating this map. This second contribution is not contained in the map calculated via Eq.~\eqref{eq:WF} since the data are not informative enough to determine the locations of all features. In a very similar fashion, Eq.~\eqref{eq:eta} estimates the correction factors for the error bars also from two main contributions. The first contribution uses simply the difference between the observed data and the data expected from the reconstructed map and the second contribution compensates partly for the attraction the data exhibit onto the map in the reconstruction step which lets some fraction of the noise imprint itself onto the map. Both contributions are rescaled by the inverse noise variance to turn this estimate of the noise variance into a correction factor. There is a third term in Eq.~\eqref{eq:eta} that is solely due to the prior we chose for $\eta$. It prevents the error bars from vanishing in case a data point is by chance in perfect agreement with the map. For a detailed derivation of these formulas, the reader is referred to \citet{oppermann-2011b}.

We include a smoothing step for the angular power spectrum in each step of the iteration, where we smooth with a kernel with $\Delta_\ell=8$ FWHM, lowering $\Delta_\ell$ for the lowest $\ell$-modes. This is done to avoid a possible perception threshold on scales with little power in the data \citep[see][]{ensslin_frommert-2011}. The smoothing step is also justified by the fact that none of the underlying physical fields, i.e.\ the thermal electron density and the line of sight component of the magnetic field, are expected to have vastly different power on neighboring scales.

\section{Data sets}
\label{sec:datasets}

\begin{figure}
	\includegraphics[width=\columnwidth]{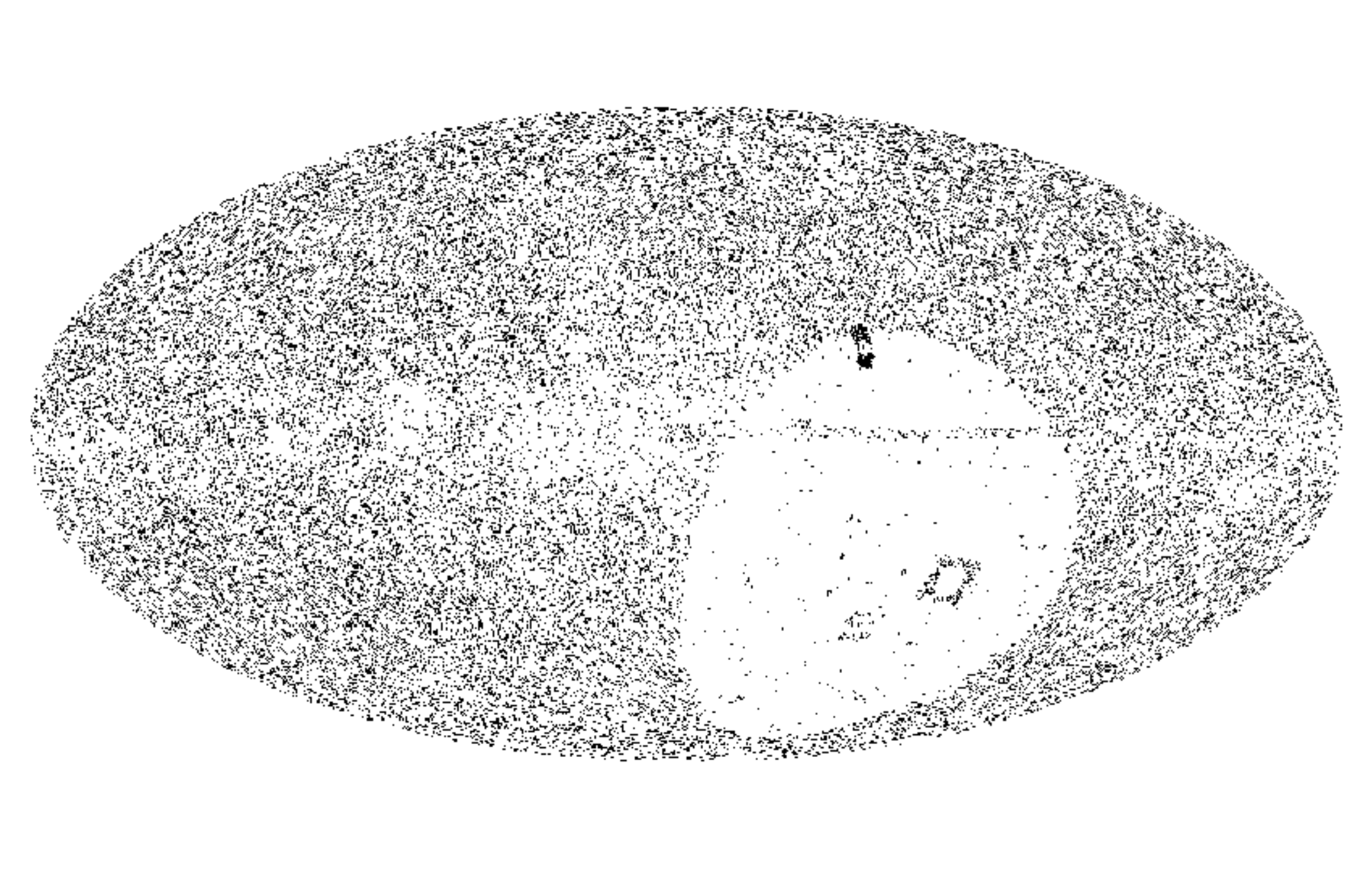}
	\caption{\label{fig:datadist}Distribution of the data points on the sky. Shown is a \textsc{HEALPix} map at a resolution of $N_\mathrm{side}=128$, using Galactic coordinates. The map is centered on the Galactic center, latitudes increase upward, and longitudes increase to the left. Each black pixel contains at least one data point.}
\end{figure}

Table \ref{tab:datasets} summarizes the data catalogs that we use for the reconstruction. Altogether, the catalogs contain 41\,330 measurements of the Faraday rotation of extragalactic point sources. Fig.~\ref{fig:datadist} shows their distribution on the sky. The coverage is clearly far from complete, especially at declinations below $-40^\circ$ where the Taylor-catalog does not provide any data. However, 24\% of the data points from the other catalogs lie within this region, so some toeholds are present even there. The densely sampled region that stands out at the top of the empty patch in Fig.~\ref{fig:datadist} is Centaurus A, studied in the Feain-catalog. The relative scarcity of data points near the Galactic plane is due to numerous depolarization effects caused by nearby structures in the magneto-ionic medium, as explained by \citet{stil-2007}. We use only extragalactic sources, and not pulsar rotation measures, since this ensures that each measurement contains the full Galactic Faraday depth.

Since the regions of coverage of the different catalogs overlap some of the data points have the same underlying radio source. While this does not constitute a problem for the reconstruction algorithm, it does in principle lead to noise correlations since the intrinsic Faraday rotation of this source, which is part of the noise in our formalism, enters each of these data points in the same way. We ignore this effect in favor of a greatly simplified analysis. The combination of the response matrix and the inverse noise covariance matrix in Eq.~\eqref{eq:WF} corresponds to an inverse noise weighted averaging of all data points that fall within one sky pixel. If the error bars were only due to the intrinsic Faraday rotation of the sources, this would amount to an underestimation of the error bar by a factor $1/\sqrt{k}$ for a source that appears in $k$ different catalogs. In reality, the intrinsic Faraday rotation constitutes only a fraction of the total error budget. The effect is therefore smaller.

Some of the catalogs listed in Table~\ref{tab:datasets} are themselves compilations of earlier measurements. As a consequence, some individual observations are contained in several of the catalogs. We have removed data points where we suspect such duplications so that each observation is used only once. Note that this does not apply to different observations of the same source, as discussed above. The number of data points given in Table~\ref{tab:datasets} is the effective number of data points that we use in our analysis from the respective catalog.

Any variation of the Galactic Faraday depth within one pixel of our map can naturally not be reconstructed. Such variations on very small scales have been detected by \citet{braun-2010} for a region around $(l,b)\approx(94^\circ,-21^\circ)$. Should several sources fall within a pixel in such a region, our algorithm will yield an appropriate average value for the pixel and increase the error bars of the data points until they are consistent with this average value.

The sources studied in the Bonafede-catalog and some of the sources in the Clarke-catalog lie within or behind galaxy clusters. They are therefore expected to have an increased extragalactic contribution to their measured Faraday rotation. In order to take the cluster contribution into account, we have corrected the error bars of these points accoring to
\begin{equation}
	\sigma_{\mathrm{(corrected)}}^2=\sigma^2+\sigma_{\mathrm{(cluster)}}^2.
\end{equation}
To estimate the cluster contribution $\sigma_{\mathrm{(cluster)}}$, \citet{bonafede-2010} studied resolved background sources for which several independent RM measurements are possible. $\sigma_{\mathrm{(cluster)}}$ was then identified with the empirical value of the standard deviation of these measurements. \citet{clarke-2001} estimated the cluster contribution by comparing the RM values of sources within the cluster to those of sources behind the cluster. The Johnston-Hollit-B-catalog also contains sources associated with galaxy clusters. However, due to the low density of sources, an estimation of the cluster contribution is not possible in this case. We expect a fraction of the other sources to be affected by clusters as well. However, since information on which sources exactly are affected is missing in general, we leave it to our algorithm to increase the error bars of the appropriate data points. The same problem exists in principle for satellite galaxies of the Milky Way, such as the Large and Small Magellanic Clouds. We do not attempt to separate their contribution to the Faraday depth from the one of the Milky Way, so that the map we reconstruct is strictly speaking not a pure map of the Galactic Faraday depth, but rather a map of the Faraday depth of the Milky Way and its surroundings. Due to our use of spatial correlations in the reconstruction algorithm, the Faraday depth contribution intrinsic to the sources will, however, be largely removed.

\begin{figure*}
	\centering
	\includegraphics[width=17cm]{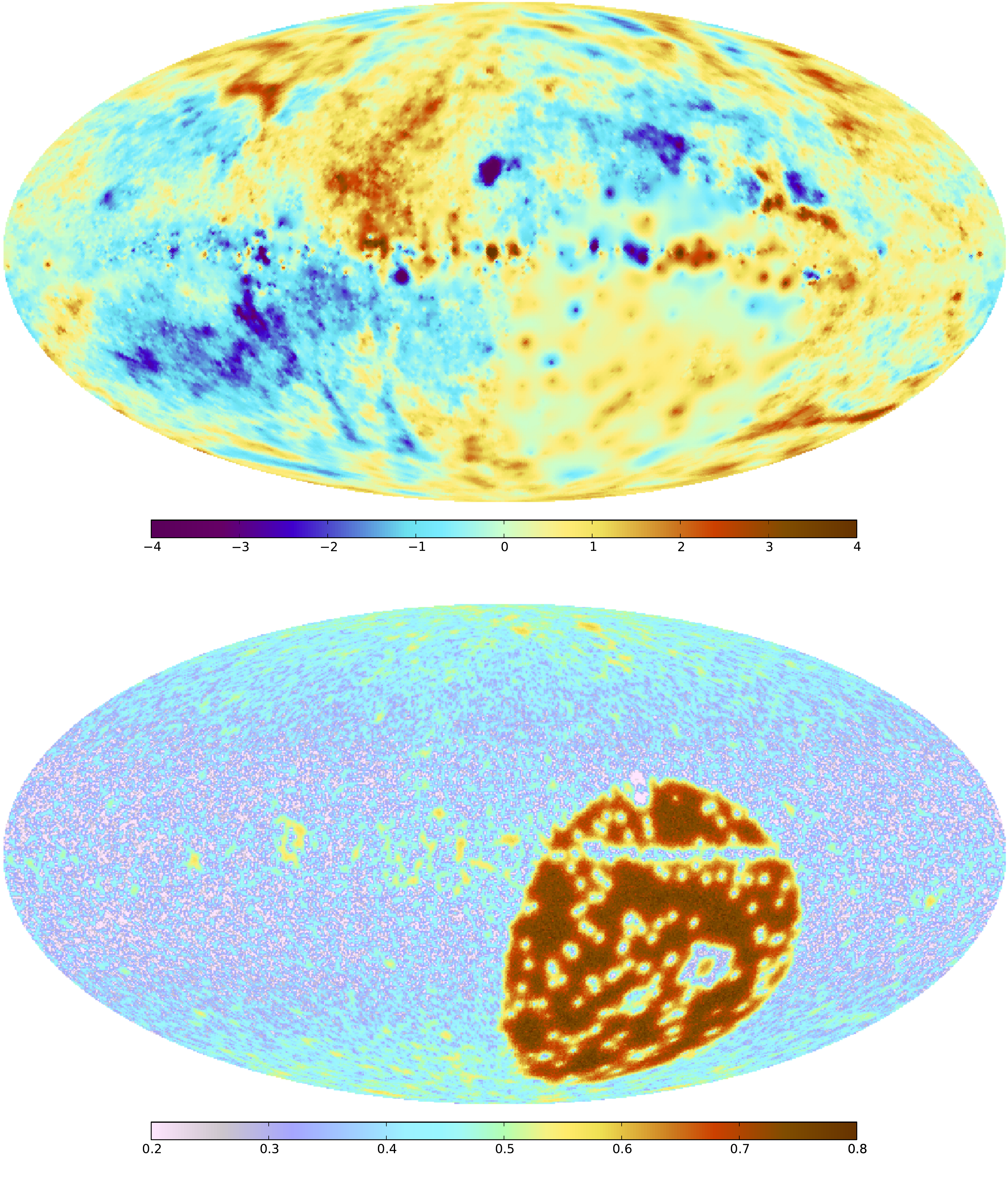}
	\caption{\label{fig:map}Reconstructed dimensionless signal map $m$ (top) and its uncertainty $\hat{D}^{1/2}$ (bottom). Note the different color codes.}
\end{figure*}

\begin{figure*}
	\centering
	\includegraphics[width=17cm]{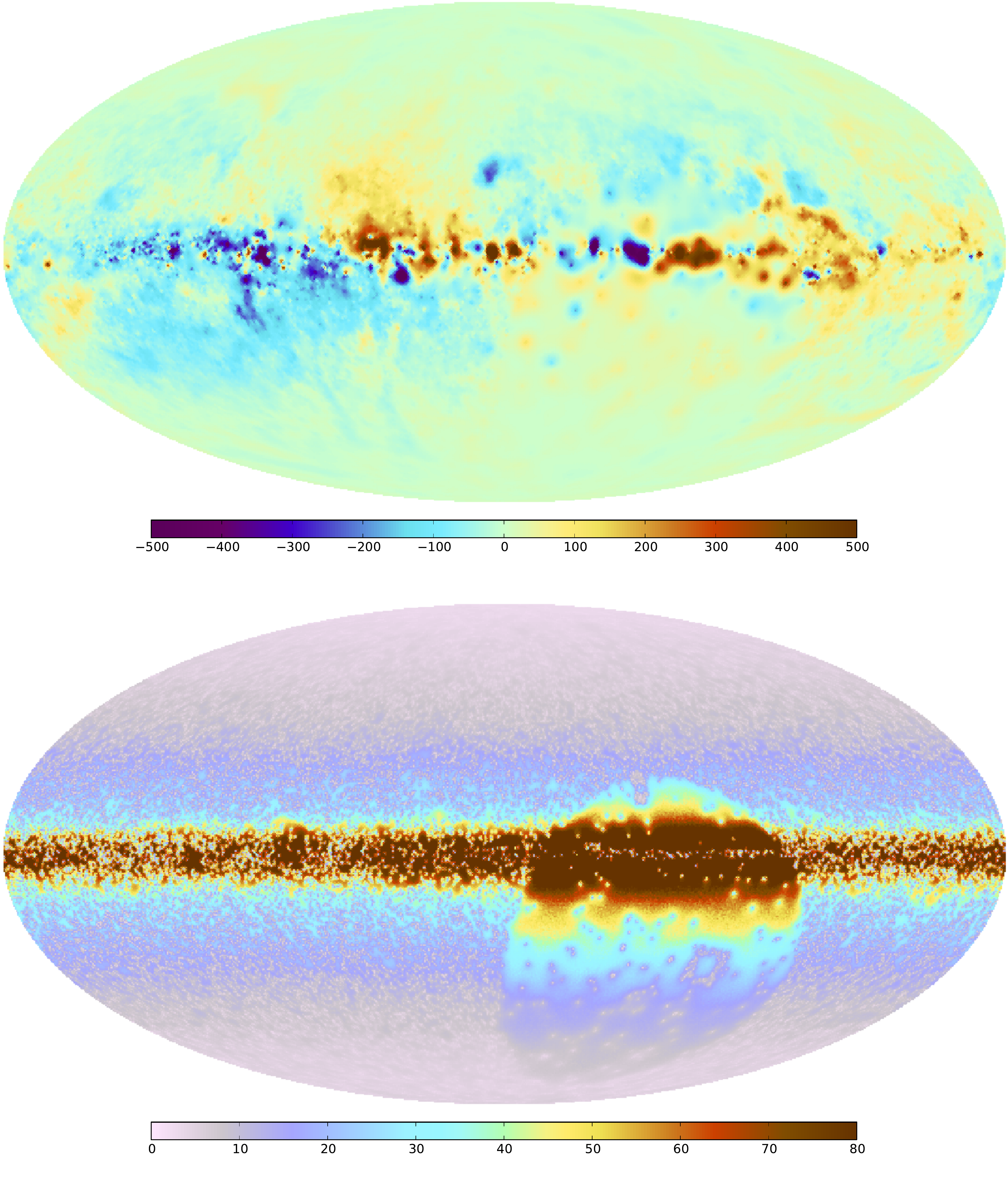}
	\caption{\label{fig:phi}Reconstruction of the Galactic Faraday depth $pm$ (top) and its uncertainty $p\hat{D}^{1/2}$ (bottom) in $\mathrm{rad}/\mathrm{m}^{2}$. Note the different color codes.}
\end{figure*}

Furthermore, some of the sources will have a non-trivial Faraday spectrum, i.e.\ they exhibit polarized emission at more than one Faraday depth. While the technique of RM synthesis \citep{burn-1966, brentjens-2005} is able to make out these sources, such features are not described by a $\lambda^2$-fit, which may thus lead to an erroneous rotation measure value. This problem becomes more severe if the number of frequencies used in the fit is low. In the limit of two frequencies, multi-component Faraday spectra necessarily go unnoticed. We use the data points obtained by $\lambda^2$-fits of only a few frequencies nevertheless, and leave it to the reconstruction algorithm to increase the error bars of those with an underlying multi-component spectrum accordingly.

\section{Results}
\label{sec:results}

All results shown here are calculated at a \textsc{HEALPix}\footnote{The \textsc{HEALPix} package is available from \url{http://healpix.jpl.nasa.gov}.} resolution of $N_\mathrm{side}=128$, i.e.\ the maps contain 196\,608 pixels. The minimum angular scale that we consider is $\ell_\mathrm{max}=383$, corresponding roughly to half a degree. These results are publicly available and can be downloaded from \url{http://www.mpa-garching.mpg.de/ift/faraday/}. The maps that we show are all centered on the Galactic center with positive Galactic latitudes at the top and positive Galactic longitudes plotted to the left.

\subsection{Map}

Figure \ref{fig:map} shows the reconstructed dimensionless signal map $m$ and an estimate for its uncertainty, given by $\hat{D}^{1/2}$. The same for the physical Galactic Faraday depth $pm$, i.e.\ the signal multiplied by the Galactic variance profile, is shown in Fig.~\ref{fig:phi}. As expected, the signal reconstruction is more uncertain in regions that lack data. Furthermore, the uncertainty in Fig.~\ref{fig:map} tends to be smaller in the Galactic plane. This is due to the higher signal response brought along by the Galactic variance profile in this area. When considering the uncertainty of the final map of the Faraday depth, i.e.\ the bottom panel of Fig.~\ref{fig:phi}, this feature gets turned around. The values within the Galactic plane now tend to be more uncertain than the ones near the poles. Note, however, that this is the absolute uncertainty. Since the Galactic Faraday depths are greater for lines of sight through the Galactic disk as well, the relative uncertainty is smaller there. This corresponds roughly to the uncertainty shown in the bottom panel of Fig.~\ref{fig:map} which can be interpreted as the uncertainty of the Galactic Faraday depth relative to the value of the Galactic variance profile at the specific latitude. Also, the uncertainty is only high in the Galactic plane in pixels that do not contain any data. In the pixels that contain measurements, the uncertainty is comparable to the error bars of the data. It should be noted, however, that due to the approximations made in the derivation of the filter formulas \citep[for details, see][]{oppermann-2011b}, the presented 1$\sigma$ intervals cannot be interpreted as containing $68\%$ of the correct pixel values of the signal. \citet{oppermann-2011b} found in their mock tests, that about $50\%$ of the correct pixel values lie within this range.

In general, Fig.~\ref{fig:map} is better suited to make out localized features away from the Galactic plane. The most striking of these features is the quadrupole-like structure on large scales that favors positive Faraday depths in the upper left and lower right quadrant and negative Faraday depths in the upper right and lower left quadrant. This has been observed in measurements of Faraday rotation in the past, first by \citet{simard-normandin-1980}, and has often been claimed to be due to a toroidal component of the large scale Galactic magnetic field that changes sign over the Galactic plane \citep[see e.g][]{han-1997}. Recent studies by \citet{wolleben-2010} and \citet{mao-2010} have shown, however, that this pattern is probably at least partly due to local features of the interstellar medium in the solar neighborhood. At Galactic longitudes beyond roughly $\pm100^\circ$, this pattern turns into a dipolar structure, favoring negative values at the left edge of the map and positive ones on the very right, as noted previously by \citet{kronberg-2011}. This might be a signature of a toroidal magnetic field component that does not change sign over the Galactic plane. But of course this could also be a local effect, independent of the large scale magnetic field.

\begin{figure}
	\centering
	\includegraphics[width=\columnwidth]{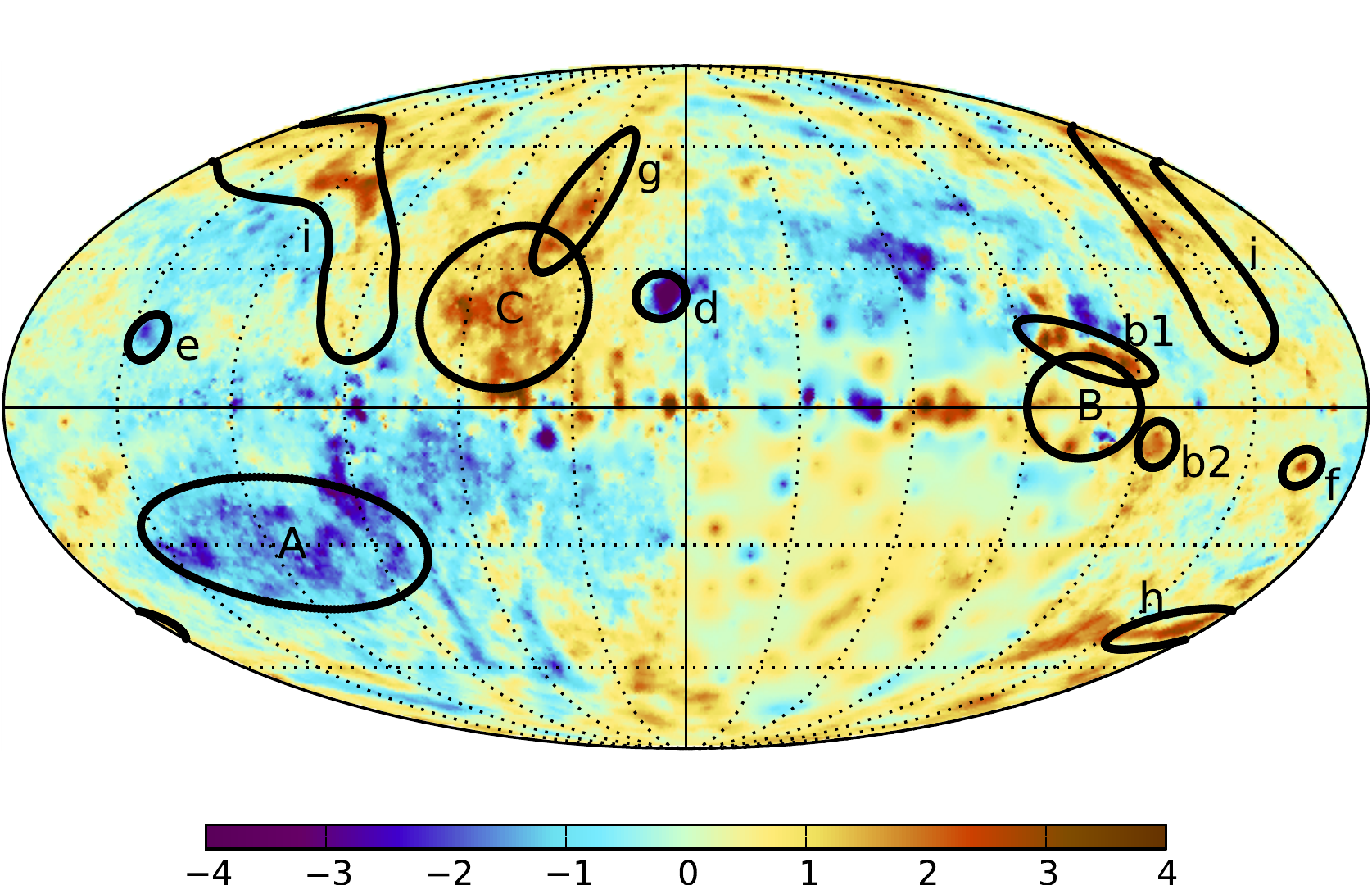}
	\caption{\label{fig:annotated}Same as the top panel of Fig.~\ref{fig:map}, with markings around the regions discussed in the text. The letters labeling the regions are used for reference in the main text. Dashed lines denote lines of constant Galactic longitude or latitude. Their angular separation is $30^\circ$.}
\end{figure}

Many other features are visible in the top panel of Fig.~\ref{fig:map}. We have marked some of the features that have already been discussed in the literature in Fig.~\ref{fig:annotated} for easier reference.

\citet{simard-normandin-1980} identified three large regions (A, B, and C in Fig.~\ref{fig:annotated}) with large angular size that stand out in Galactic Faraday depth amplitude. \citet{stil-2011} narrowed the definitions of the regions A and C down to their more striking parts using the NVSS RM catalog. Region A is a large area of negative Galactic Faraday depth localized roughly at $80\degr < l < 150\degr$, $-40\degr < b < -20\degr$. This region is seen in the direction of radio Loop II, but there is little evidence that the two are associated. The high-longitude boundary of region A coincides with part of the edge of Loop II. However, pulsar rotation measures suggest that Region A extends more than 3 kpc along the line of sight \citep{simard-normandin-1980}, which suggests that region A is a much larger structure.

Region B of \citet{simard-normandin-1980} is associated with the Gum nebula. \citet{vallee-1983b} and \citet{stil-2007} identified a large magnetic shell in the area. The arc of positive Galactic Faraday depth around $250~\mathrm{rad}/\mathrm{m}^2$ at $-120\degr < l < -90\degr$, $b \approx 13^\circ$ (region b1 in Fig.~\ref{fig:annotated}) coincides with the northern H$\alpha$ arc of the Gum nebula. A small excess in Galactic Faraday depth (region b2 in Fig.~\ref{fig:annotated}) is associated with the nearby HII region RCW 15 ($l = -125^\circ$, $b = -7^\circ$).

Region C is an area of positive Galactic Faraday depth in the range $33\degr < l < 68^\circ$, $10\degr < b < 35^\circ$ near the boundary of Radio Loop I. \citet{wolleben-2010} found diffuse polarized emission at a Faraday depth of $60~\mathrm{rad}/\mathrm{m}^2$ at $l \approx 40^\circ$, $b \approx 30^\circ$ with associated HI structure, and interpreted this structure as part of a separate super shell around a subgroup of the Sco-Cen (Sco OB2\_2).

Besides the Gum nebula, some extended HII regions at intermediate Galactic latitude can be identified in the form of a localized excess in Galactic Faraday depth \citep{stil-2007, harvey-smith-2011}. The HII regions Sh 2-27 around $\zeta$ Oph at $l= 8^\circ, b = 23.5^\circ$ (region d in Fig.~\ref{fig:annotated}) and Sivan 3 around $\alpha$ Cam at $l = 144.5^\circ$, $b = 14^\circ$ (region e in Fig.~\ref{fig:annotated}) stand out as isolated regions of negative Galactic Faraday depth, while Sh 2-264 around $\lambda$ Ori (region f in Fig.~\ref{fig:annotated}) is visible as a positive excess at $l = 195$, $b = -12$. \citet{stil-2011} presented an image of H$\alpha$ intensity with rotation measure data overplotted.

Some large shells are also visible in the image of the Galactic Faraday depth. The Galactic anti-center direction is the most favourable direction to see these large structures, because it is less crowded than the inner Galaxy and the line of sight makes a large angle with the large-scale magnetic field. The North Polar Spur (region g in Fig.~\ref{fig:annotated}) is the notable exception toward the inner Galaxy. The filament of positive Galactic Faraday depth at $180^\circ < l < 200^\circ$, $b \approx -50^\circ$ (region h in Fig.~\ref{fig:annotated}) is associated with the wall of the Orion-Eridanus superbubble \citep{heiles-1976, brown-1995}. A large arc of positive Galactic Faraday depth (region i in Fig.~\ref{fig:annotated}) rises north of the Galactic plane at around $l \approx 95^\circ$ up to $b \approx 65^\circ$ around $l = 180^\circ$ and curves back to the Galactic plane at around $l = 210^\circ$ \citep{stil-2011}. This arc of positive Galactic Faraday depth traces the intermediate-velocity arch of atomic hydrogen gas identified by \citep{kuntz-1996}.

\citet{xu-2006} reported RM excesses in the direction of the nearby Perseus-Pisces and Hercules super clusters. The higher sampling provided by the new Faraday rotation data catalogs has revealed high-latitude structures in the Galactic Faraday depth that warrant further investigation in the effect of the Galactic foreground. Many more small- and intermediate-scale features are visible in the top panel of Fig.~\ref{fig:map}. A detailed analysis of these features is left for future work.

\subsection{Reconstruction of the noise covariance}
\label{sec:noise-rec}

The extended critical filter adapts the correction factors $(\eta_i)_i$, introduced in Sect.~\ref{sec:algorithm}, so as to make the error bars of the data conform with the local map reconstruction. This is influenced by the surrounding data points and the angular power spectrum, which is in turn reconstructed using the entirety of the data. \citet{oppermann-2011b} showed that allowing for this adaptation of the error bars leads to a slight oversmoothing of the reconstructed map since small-scale features that are only supported by individual data points get easily misinterpreted as noise.

In our reconstruction, we find that the median correction factor is $\eta^\mathrm{(med)}=0.56$. This indicates that the bulk of the data points are rather consistent with one another and therefore with the reconstruction as well. As a consequence, their error bars are not enlarged but rather slightly decreased by the algorithm. Oversmoothing can therefore not be a serious issue for the map as a whole. This is supported by the geometric mean of the correction factors, for which we find $\eta^{\mathrm{(geom)}}=0.75$. This corresponds to the arithmetic mean on a logarithmic scale and its prior expectation value was tuned to be one. Looking at the arithmetic mean on a linear scale, we find $\eta^\mathrm{(mean)}=6.40$, indicating that there are at least a few data points whose error bars get corrected upward significantly. In fact, there are $134$ data points with $\eta_i>400$, meaning that the error bar has been increased by a factor of more than $20$. These are isolated outliers in the data that are not consistent with their surroundings.

\begin{figure}
	\input{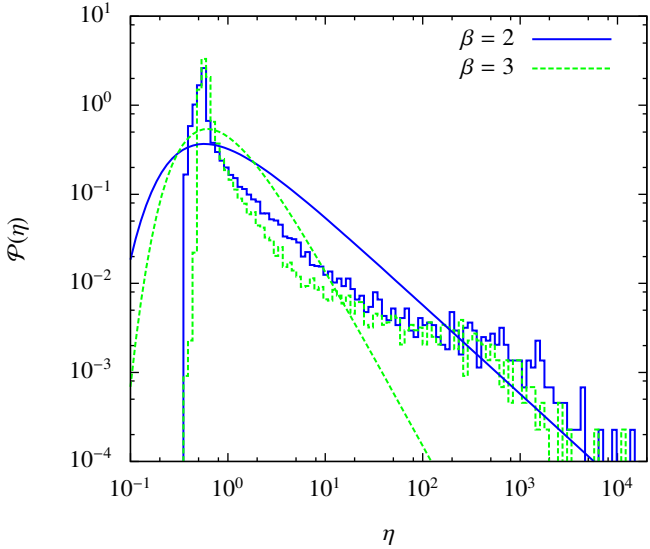}
	\caption{\label{fig:hist}Comparison of the reconstructed distribution of the correction factors $\eta$ that enter the noise covariance matrix and their priors. The dark histogram and line show the normalized distribution and prior for $\beta=2$, respectively. The light histogram and line show the same for $\beta=3$.}
\end{figure}

\begin{figure}
	\input{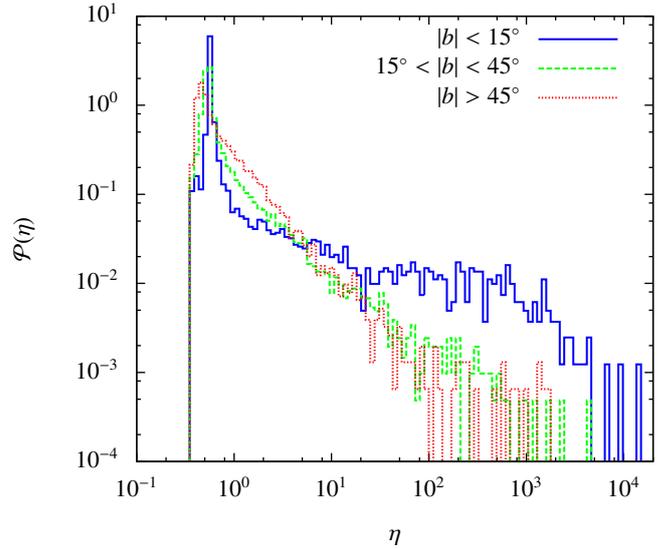}
	\caption{\label{fig:hist_latitude}Comparison of the reconstructed distributions of the correction factors $\eta$ for different latitude bins. The dark solid histogram depicts the distribution for data points within the Galactic plane, the light dashed histogram the distribution for data points at intermediate latitudes, and the dotted histogram the one for data points in the polar regions. Only the results obtained with $\beta=2$ are shown.}
\end{figure}

\begin{figure}
	\input{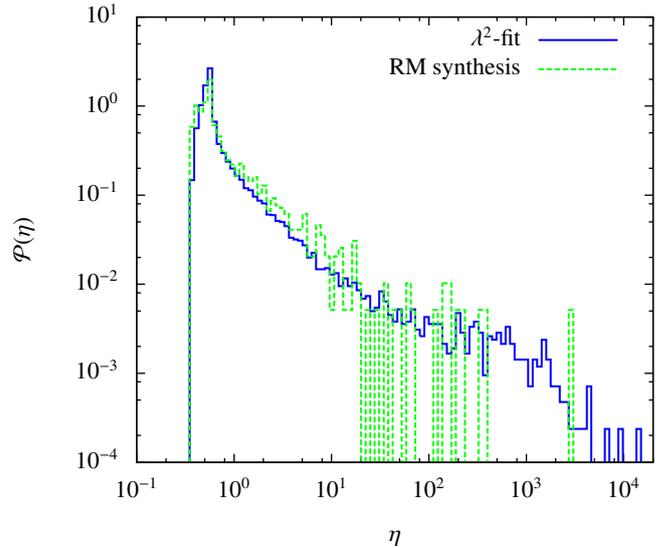}
	\caption{\label{fig:hist_method}Comparison of the reconstructed distributions of the correction factors $\eta$ for the two data reduction techniques. The dark solid histogram depicts the distribution for data points obtained from a linear $\lambda^2$-fit and the light dashed histogram the distribution for data points stemming from RM synthesis studies.}
\end{figure}

Figure~\ref{fig:hist} shows the final distribution of $\eta$-values. The bulk of these values lie around $\eta=1$ or even slightly below. Only relatively few data points have highly increased error bars (note the logarithmic scale of the vertical axis in Fig.~\ref{fig:hist}). Also plotted in Fig.~\ref{fig:hist} is the distribution of $\eta$-values that resulted from a reconstruction in which the slope parameter in the prior for the correction factors was chosen to be $\beta=3$, as well as the prior probability distributions corresponding to $\beta=2$ and $\beta=3$. This shows two things. The resulting distribution does not change much when the value of $\beta$ is changed and both distributions are better represented by a prior with $\beta=2$. Our choice for $\beta$ is thus justified.

The data points with $\eta\gg1$ do not appear to be spatially clumped, making it improbable that any extended physical features that are present in the data are lost due to the increase in the assumed noise covariance. Any real features that might mistakenly be filtered out in this procedure can be expected to be smaller or comparable in size to the distance to the next data point, i.e.\ one or two pixels or about one degree in most parts of the sky. The data points with strongly corrected error bars are predominantly located near the Galactic plane. This can be clearly seen in Fig.~\ref{fig:hist_latitude}, where we plot the distribution of the correction factors for three latitude bins separately. While the difference in the distributions for the polar regions and the intermediate latitude bin is not very big, the data points around the Galactic disk clearly are more likely to have correction factors at the high end. At least in some cases these high $\eta$-values can be interpreted as correcting an offset in the rotation angle of $\pi$ that has escaped the observational analysis. Others might be due to a high level of polarized emissivity within the Galactic disk that can lead to misleading RM fits. Another reason for high $\eta$-values is a higher extragalactic contribution to the measured Faraday rotation, caused e.g. by magnetic fields in galaxy clusters. This last reason, however, would not be expected to show any statistical latitude dependence.

As mentioned earlier, a non-trivial emission spectrum in Faraday space is hard to identify when using linear $\lambda^2$-fits to obtain RM values. We therefore compare the distributions of the correction factors for data points from $\lambda^2$-fits and the ones for data points that stem from RM synthesis studies in Fig.~\ref{fig:hist_method}. From the histograms it can indeed be seen that the data from $\lambda^2$-fits are more likely to have a high $\eta$-value, as expected.

\begin{figure}
	\includegraphics[width=\columnwidth]{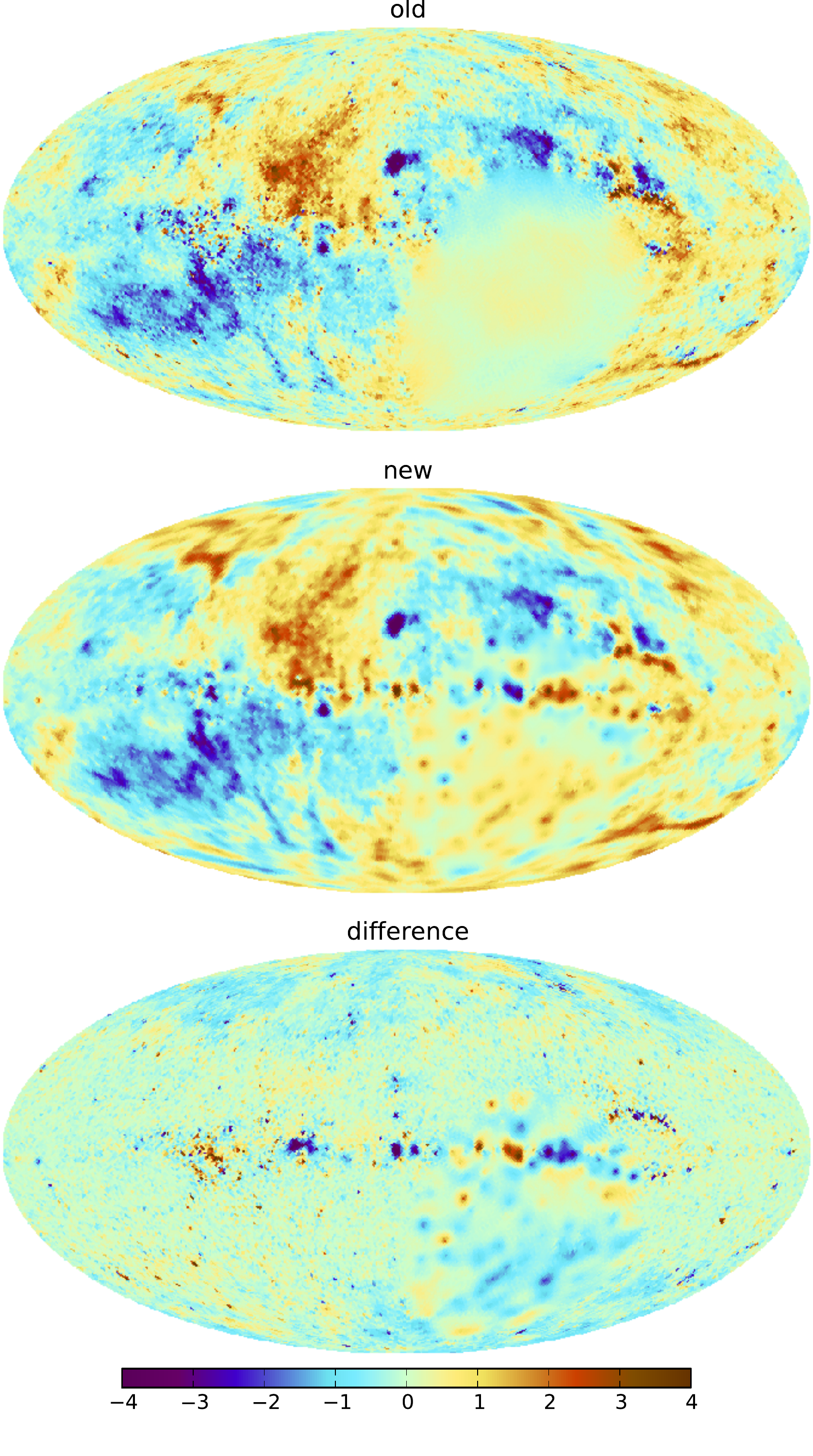}
	\caption{\label{fig:oldcomp}Comparison of the reconstruction of the dimensionless signal to earlier results. The top panel shows the reconstructed signal field of \citet{oppermann-2011a}, the middle panel shows the same as the top panel of Fig.~\ref{fig:map}, only coarsened to a resolution of $N_\mathrm{side}=64$ to match the resolution of the old reconstruction. The bottom panel shows the difference between the upper panel and the middle panel.}
\end{figure}

Figure~\ref{fig:oldcomp} shows a comparison of our reconstructed signal map with the reconstruction of \citet{oppermann-2011a}, where the critical filter formalism was used without accounting for uncertainties in the noise covariance and only data from the Taylor-catalog were used. The differences that can be seen are twofold. On the one hand, our map shows structure due to the additional data points that we use, most prominently at declinations below $-40^\circ$. On the other hand, some of the features present in the older map have vanished since they were supported only by a single data point which has been interpreted as being noise-dominated by our algorithm. These features appear prominently both in the old map and in the difference map, where our newly reconstructed map has been subtracted from the old one. They have the same sign in both these maps. Also, our new reconstruction is less grainy. This is a combined effect of the higher resolution that we use and the adaptation of error bars during our reconstruction.

\subsection{Power spectrum}

\begin{figure}
	\input{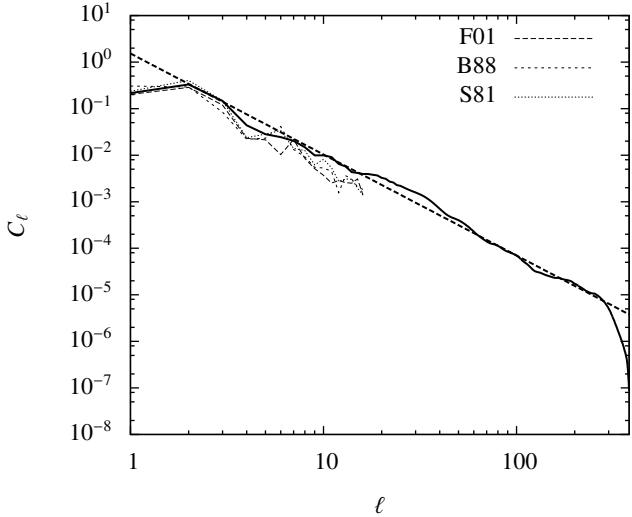}
	\caption{\label{fig:angspec}Angular power spectrum of the dimensionless signal field (thick solid line), along with a power law fit, $C_\ell\propto\ell^{-2.17}$ (thick dashed line). The thin lines depict the angular power spectra corresponding to the maps reconstructed by \citet{dineen-2005}, corrected for the Galactic variance profile. The three RM catalogs used in their work are from \citet{simard-normandin-1981} (S81), \citet{broten-1988} (B88), and \citet{frick-2001} (F01).}
\end{figure}

\begin{figure}
	\input{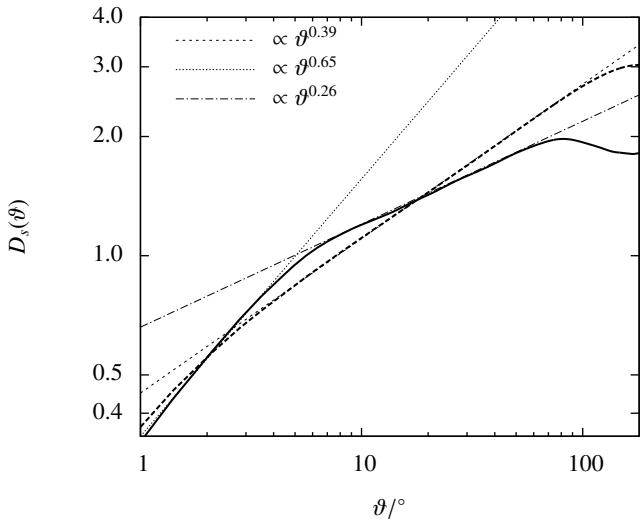}
	\caption{\label{fig:structurefunction}Second order structure function corresponding to the angular power spectrum plotted in Fig.~\ref{fig:angspec} (thick solid line) and its power law fit (thick dashed line), along with power law approximations (thin lines).}
\end{figure}

The reconstructed angular power spectrum of the dimensionless signal field is shown in Fig.~\ref{fig:angspec}. It is well described by a power law. A logarithmic least square fit, which is also shown in Fig.~\ref{fig:angspec}, yields a spectral index of $2.17$, i.e.
\begin{equation}
	C_\ell\propto\ell^{-2.17},
        \label{eq:Clfit}
\end{equation}
where we have taken scales down to $\ell=300$ into account. Note that due to the typical distance of neighboring data points of roughly one degree, structures smaller than this angular size, corresponding to $\ell\gtrsim180$, will in general not be reconstructed and we might therefore be missing some power on the smallest scales. However, some data points have smaller angular separations and we therefore have some information on the angular power spectrum up to $\ell_\mathrm{max}=383$.

Also shown in Fig.~\ref{fig:angspec} is a comparison with the angular power spectra of the maps that \citet{dineen-2005} reconstructed. They created three separate maps from three different RM catalogs. We used the spherical harmonics components of their maps\footnote{\citet{dineen-2005} provide their results at \url{http://astro.ic.ac.uk/~pdineen/rm_maps/}.}, transformed them to position space, and then divided them by our Galactic variance profile. We plot the angular power spectra of the three resulting dimensionless maps. Evidently, both the slope and the normalization of the spectra are in agreement with our result. \citet{haverkorn-2003} study the angular power spectrum of rotation measures of diffuse polarized radio emission from the local interstellar medium in two regions of the sky on scales $400<\ell<1500$. They fit power laws with exponents close to $-1$, i.e.\ $C_\ell\propto\ell^{-1}$, significantly larger than our result. This is not necessarily a contradiction, however, since a flattening of the angular power spectrum on scales that are too small for our analysis could explain both results. Furthermore, we take into account the full line of sight through the galaxy by using only extragalactic sources, so the volume that we probe is significantly larger than the one probed by \citet{haverkorn-2003}.

In order to compare our result to other earlier papers, we consider the second order structure function for the dimensionless signal field,
\begin{align}
	\label{eq:structurefunction}
	D_s(\vartheta)&=\left<\left(s(\hat{n})-s(\hat{n}')\right)^2\right>_{\mathcal{P}(s)}\nonumber\\
	&=2\left(S_{\hat{n}\hat{n}}-S_{\hat{n}\hat{n}'}\right),
\end{align}
where $\vartheta=\arccos(\hat{n}\cdot\hat{n}')$ and $\hat{n}$ and $\hat{n}'$ are two directions in the sky. Here, $S$ denotes the signal covariance matrix and the angle brackets denote a prior ensemble average. Since we assume statistical homogeneity and isotropy for the signal field, $S_{\hat{n}\hat{n}}$ does not depend on $\hat{n}$, $S_{\hat{n}\hat{n}'}$ depends only on $\vartheta$, and both terms are completely determined by the angular power spectrum. This also allows us to exchange the usual spatial average with an ensemble average in Eq.~\eqref{eq:structurefunction}. The resulting structure function is plotted in Fig.~\ref{fig:structurefunction}. Using the final angular power spectrum of our reconstruction (the solid line in Fig.~\ref{fig:angspec}), we find a broken power law with exponents $0.65$ for small angles and $0.26$ for large angles with the transition occuring around $\vartheta=5^\circ$ (the solid line in Fig.~\ref{fig:structurefunction}). The power law fit to the angular power spectrum (the dashed line in Fig.~\ref{fig:angspec}) leads to a structure function that can be approximated by a single power law with exponent $0.39$ (the dashed line in Fig.~\ref{fig:structurefunction}).

\citet{minter-1996} found that the structure function derived from their observations is well described by a power law with exponent $0.64$ for angular scales of $\vartheta>1^\circ$. \citet{sun-2004} study the structure function in three different regions within the Galactic plane and in the vicinity of the North Galactic pole. An inverse noise weighted average of their power law indices yields a value of $0.11$. \citet{haverkorn-2006a} and \citet{haverkorn-2008} study observations through interarm regions in the Galactic plane separately from observations along Galactic arms. They find flat structure functions for the observations along Galactic arms. \citet{haverkorn-2006a} find a weighted mean power law index of $0.55$ for the structure functions derived from observations through interarm regions, while \citet{haverkorn-2008} find an inverse-noise weighted mean power law index of $0.40$. \citet{haverkorn-2003} find flat structure functions for the two regions that they study. \citet{roy-2008} find a structure function for the region around the Galactic center that is constant on scales above $\vartheta=0.7^\circ$ and exhibits a power law behavior with an exponent of $0.7$ on smaller scales. \citet{stil-2011} fit broken power laws with the breaking point at $\vartheta=1^\circ$ to the structure functions they extract from the NVSS rotation measure catalog \citep{taylor-2009}. They find power law indices that vary spatially. Taking an inverse-noise weighted average of their power law indices for the regions that they study in detail yields $0.37$ for $\vartheta>1^\circ$ and $0.59$ for $\vartheta<1^\circ$.

These observational results indicate that the slope of the structure function varies from region to region. Our result is insensitive to these variations since our structure function is just a description of the prior for the dimensionless signal, for which we have assumed statistical isotropy. It can therefore be interpreted as a mean structure function across the whole sky. The observations that yield non-flat structure functions are in rough agreement with the slopes that we fit in Fig.~\ref{fig:structurefunction}. The dependence of the structure function slope on Galactic latitude \citep[e.g.][]{simonetti-1984, sun-2004} is partly removed in our analysis by the division through the Galactic variance profile. Note that \citet{simonetti-1984, simonetti-1986} already suspected a break in the structure function at roughly five degrees. However, existing studies have not shown convincing evidence for this.

\subsubsection{Consequences for the 3D fields}

As an illustrative thought experiment, assume that an observer is sitting in the middle of a spherical distribution of magnetoionic medium. Let $\tilde{\varphi}(\vec{x})\propto n_\mathrm{e}(\vec{x})B_r(\vec{x})$ be the product of the local thermal electron density and the line of sight component of the magnetic field as a function of 3D position $\vec{x}$, i.e.\ the differential contribution to the Faraday depth that this observer is measuring. We model this field as factorizing into two parts,
\begin{equation}
	\tilde{\varphi}(\vec{x})=\bar{\varphi}(r)\varphi(\vec{x}).
\end{equation}
The first part is a spherically symmetric contribution, whose functional dependence on the radial distance from the observer is known, and the second part is assumed to be a realization of a statistically homogeneous and isotropic random field, i.e.
\begin{equation}
	\label{eq:phistatistics}
	\left<\varphi(\vec{k})\varphi^*(\vec{k}')\right>=\left(2\pi\right)^3\delta^{(3)}(\vec{k}-\vec{k}')P_\varphi(k),
\end{equation}
where the angle-brackets denote an average over all possible field realizations, $P_\varphi(k)$ is the Fourier power spectrum\footnote{Note that the definition of the Fourier power spectrum made in Eq.~\eqref{eq:phistatistics} corresponds to what is sometimes referred to as the 3D power spectrum, i.e.\ the variance of the field $\varphi$ at each position $\vec{x}$ in real space can be calculated as $\left<\varphi^2(\vec{x})\right>\propto\int_0^\infty\mathrm{d}k~k^2P_\varphi(k)$.} that describes the statistics of $\varphi$, and $k=\left|\vec{k}\right|$.

Using the simplest form of $\bar{\varphi}(r)$, namely a constant within some finite radius $r_0$, i.e.
\begin{equation}
	\bar{\varphi}(r)=\left\{
		\begin{array}{cc}
			\varphi_0 & \textrm{if}~r<r_0\\
			0 & \textrm{else}
		\end{array}
	\right.,
\end{equation}
and a power law for the Fourier power spectrum,
\begin{equation}
	P_\varphi(k)\propto k^{-\alpha},
\end{equation}
we calculated the angular power spectrum of the Faraday depth that the observer would measure and compared the result numerically with Eq.~\eqref{eq:Clfit}. We find that the two agree well if one chooses $\alpha$ roughly equal to the power law index that was found for the angular power spectrum, i.e.\ $2.17$ in this case.

A similar thought experiment has been conducted by \citet{simonetti-1984}. They assume a Fourier power spectrum $P_\varphi(k)\propto\exp\left(-k^2/k_1^2\right)\left(1+k^2/k_0^2\right)^{\alpha/2}$, i.e.\ a power law with a low-wavenumber cutoff at $k_0$ and a high-wavenumber cutoff at $k_1$, and calculate the expected structure function. In the power law regime, i.e.\ $1/k_1\ll r_0\sin\vartheta\ll 1/k_0$, they find $D_s(\vartheta)\propto\vartheta^{\alpha-2}$ to lowest order in $\vartheta$. Extending this study to independent variations in the thermal electron density and the magnetic field component along the line of sight, each described by a power law power spectrum with the same index $\alpha$, \citet{minter-1996} found the same dependence on $\vartheta$.\footnote{\citet{minter-1996} assume a rectangular shape for $\varphi_0$ instead of a spherical one.} Our intermediate fit of $D_s(\vartheta)\propto\vartheta^{0.39}$ (see Fig.~\ref{fig:structurefunction}) therefore corresponds to $\alpha=2.39$, in rough agreement with our numerical finding from the power spectrum analysis.

\citet{armstrong-1995} have used observations of effects of interstellar radio scintillation \citep[see also][]{rickett-1977,rickett-1990}, as well as pulsar dispersion measures, to constrain the power spectrum describing the fluctuations of the thermal electron density in the local interstellar medium. They found a Kolmogorov-type power spectrum, i.e.\ a power law index of $\alpha=11/3$ in the present notation. This result was combined by \citet{minter-1996} with their own observations of rotation measures of extragalactic sources. Since they do not find the slope expected from the Kolmogorov power law in the structure function of the rotation measure they observe, they conclude that the outer scale of the Kolmogorov-type turbulence is smaller than the smallest scale probed by their RM observations. They fit model structure functions for the variations of the thermal electron density and the magnetic field to their own observations of RM, as well as observations of $\mathrm{H}\alpha$ intensity and $\mathrm{H}\alpha$ velocity performed by \citet{reynolds-1980}, while also taking into account the results of \citet{armstrong-1995} on smaller scales. This procedure leads to an estimate of the angular scale corresponding to the outer scale of the turbulence in the region of their observations of $\vartheta^{\mathrm{(out)}}\lesssim0.1^\circ$. Although the outer scale of the turbulence may well vary across the Galaxy, it is probably safe to assume that the scales larger than one degree that are mainly probed by the observations used in this work, are not dominated by three-dimensional turbulence. Whether or not the simple power law behavior of the angular power spectrum in Eq.~\eqref{eq:Clfit} points to some sort of interaction between the fluctuations on different scales is at the moment an open question.

In any case it is clear that the simplifying assumptions made in the thought experiments presented above are far from the truth in the Galactic setting. A more realistic study will likely have to involve numerical magneto-hydrodynamical simulations of the interstellar medium, which have become more and more sophisticated over the last years \citep[see e.g.][]{avillez-2007, kissmann-2008, burkhart-2009, tofflemire-2011}. Cross-checking the angular power spectrum of the Faraday depth that is predicted by such a simulation against Eq.~\eqref{eq:Clfit} might be a good indicator of how realistic the simulation actually is. For this, an empiric variance profile would have to be calculated from the simulated observations to create a dimensionless signal field comparable to our reconstruction. Numerical studies will also be able to show whether the simple power law that we find for the angular power spectrum is a functional form that arises generically or an outcome that needs certain ingredients. This may then enable a physical interpretation of the angular power spectrum that we find. On the other hand, if simulations show that different physical processes are needed to create the fluctuation power on different angular scales, our result will directly constrain the relative strength of these processes.

\section{Conclusions}
\label{sec:conclusions}

We have presented a map of the Galactic Faraday depth that summarizes the current state of knowledge, along with its uncertainty. For the map reconstruction we have used the extended critical filter, a state-of-the-art algorithm, yielding a result that is robust against individual faulty measurements. It is this robustness, along with the usage of the most complete data set on the Faraday rotation of extragalactic sources to date, and the high resolution that we are therefore able to reach, that make our map an improvement over existing studies. Along with the map, the reconstruction algorithm yields the angular power spectrum of the underlying signal field, $C_\ell\propto\ell^{-2.17}$, which is in agreement with earlier work. We have discussed the implications of this power spectrum for the statistics of the 3D quantities involved in a greatly simplified scenario and suggested future work on simulations with the possibility of checking predicted angular power spectra against our observational result.

All products of this work, i.e.\ the maps and their uncertainties, as well as the angular power spectrum, are made available to the community\footnote{See \url{http://www.mpa-garching.mpg.de/ift/faraday/} for a fits-file containing all the results and an interactive map to explore the Galactic Faraday sky.} for further analysis, interpretation, and for use in other work where the Galactic Faraday depth plays a role.

\begin{acknowledgements}
The authors would like to thank Steven R. Spangler for the valuable contributions he made to this paper as a referee. N.O. thanks Marco Selig and Maximilian Ullherr for fruitful discussions during the genesis of this work. Some of the results in this paper have been derived using the \textsc{HEALPix} \citep{gorski-2005} package. The calculations were performed using the \textsc{SAGE} \citep{sage} mathematics software. This research has made use of NASA's Astrophysics Data System. This research was performed in the framework of the DFG Forschergruppe 1254 ``Magnetisation of Interstellar and Intergalactic Media: The Prospects of Low-Frequency Radio Observations''. Basic research in radio astronomy at the Naval Research Laboratory is funded by 6.1 Base funding. B.M.G. and T.R. acknowledge the support of the Australian Research Council through grants FF0561298, FL100100114 and FS100100033. M.J.-H. and L.P. acknowledge support via Victoria University of Wellington Faculty of Science and Marsden Development Fund research grants awarded to M.J.-H. The Australia Telescope Compact Array is part of the Australia Telescope National Facility which is funded by the Commonwealth of Australia for operation as a National Facility managed by CSIRO. This paper includes archived data obtained through the Australia Telescope Online Archive\footnote{\url{http://atoa.atnf.csiro.au}}.
\end{acknowledgements}

\bibliographystyle{myaa}
\bibliography{FD_map}

\begin{landscape}
\begin{table}
	\caption{\label{tab:datasets}Details of the data sets used for the map reconstruction.}
	\centering
	\begin{tabular}{lcccccccc}
		identifyer & telescope & survey & \# observed wavelengths & frequency range / MHz & method & \# data points & catalog reference & survey reference\\
		\hline
		Bonafede & VLA & & 3-5 & various\tablefootmark{(a)} & $\lambda^2$-fit & 7 & (1) & \\
		Broten & various\tablefootmark{(b)} & & various\tablefootmark{(b)} & various\tablefootmark{(b)} & $\lambda^2$-fit & 121+3/2\tablefootmark{(c)} & (2) & \\
		Brown CGPS & DRAO ST & CGPS & 4 & 1\,403-1\,438 & $\lambda^2$-fit & 380 & (3) & (4)\\
		Brown SGPS & ATCA & SGPS & 12 & 1\,332-1\,436 & $\lambda^2$-fit & 148 & (5) & (6),(7)\\
		Clarke & VLA & & 4,6 & 1\,365-4\,885 & $\lambda^2$-fit & 125 & (8),(9) & \\
		Clegg & VLA & & 6 & 1\,379-1\,671 & $\lambda^2$-fit & 56 & (10) & \\
		Feain & ATCA & Cent.~A & 24 & 1\,280-1\,496 & RM synthesis & 281 & (11) & (12)\\
		Gaensler & ATCA & SGPS test & 9 & 1\,334-1\,430 & $\lambda^2$-fit & 18 & (13) & \\
		Hammond & ATCA & & 23 & 1\,332-1\,524 & RM synthesis & 88 & (14) & \\
		Heald & WSRT & WSRT-SINGS & 1024 & 1\,300-1\,763 & RM synthesis & 57 & (15) & (16)\\
		Hennessy & VLA & & 4 & 1\,362-1\,708 & $\lambda^2$-fit & 17 & (17) & \\
		Johnston-Hollitt A & ATCA & & 23 & 1\,292-1\,484 & RM synthesis & 68 & (18) & \\
		Johnston-Hollitt B & ATCA & & 4 & 1\,384-6\,176 & $\lambda^2$-fit & 12 & (19),(20) & \\
		Kato & Nobeyama & & 4\tablefootmark{(d)} & 8\,800-10\,800\tablefootmark{(d)} & $\lambda^2$-fit & 1 & (21) & \\
		Kim & various\tablefootmark{(e)} & & various\tablefootmark{(e)} & various\tablefootmark{(e)} & $\lambda^2$-fit & 20+1/2\tablefootmark{(c)} & (22) & \\
		Klein & VLA \& Effelsberg & B3/VLA & 4 & 1\,400-10\,600 & $\lambda^2$-fit & 143 & (23) & (24),(25)\\
		Lawler & various\tablefootmark{(f)} & & various\tablefootmark{(f)} & various\tablefootmark{(f)} & $\lambda^2$-fit & 3 & (26) & (27)\\
		Mao SouthCap & ATCA & & 32 & 1\,320-2\,432 & RM synthesis & 329 & (28) & \\
		Mao NorthCap & WSRT & & 16 & 1\,301-1\,793 & RM synthesis & 400 & (28) & \\
		Mao LMC & ATCA & & 14 & 1\,324-1\,436 & RM synthesis & 188 & (29),(30) & \\
		Mao SMC & ATCA & & 14 & 1\,324-1\,436 & $\lambda^2$-fit & 62 & (31) & \\
		Minter & VLA & & 4 & 1\,348-1\,651 & $\lambda^2$-fit & 98 & (32) & \\
		Oren & VLA & & 4,6 & various\tablefootmark{(g)} & $\lambda^2$-fit & 51+4/2\tablefootmark{(c)} & (33) & \\
		O'Sullivan & ATCA & & 100 & 1\,100-2\,000 & RM synthesis & 46 & (34) & \\
		Roy & ATCA \& VLA & & 4 and more & various\tablefootmark{(h)} & $\lambda^2$-fit & 67 & (35)& \\
		Rudnick & VLA & & 2 & 1\,440-1\,690 & $\lambda^2$-fit & 17+2/2\tablefootmark{(c)} & (36) & \\
		Schnitzeler & ATCA & & 12 & 1\,320-1\,1\,448\tablefootmark{(i)} & RM synthesis & 178 & (37) & \\
		Simard-Normandin & various\tablefootmark{(j)} & & various\tablefootmark{(j)} & various\tablefootmark{(j)} & $\lambda^2$-fit & 535+6/2\tablefootmark{(c)} & (38) & \\
		Tabara & various\tablefootmark{(k)} & & various\tablefootmark{(k)} & various\tablefootmark{(k)} & $\lambda^2$-fit & 62+3/2\tablefootmark{(c)} & (39) & \\
		Taylor & VLA & NVSS & 2 & 1\,344-1\,456 & $\lambda^2$-fit & 37\,543 & (40) & (41)\\
		Van Eck & VLA & & 14 & 1\,353-1\,498 & RM synthesis\tablefootmark{(l)} & 194 & (42) & \\
		Wrobel & VLA & & 6 & 1\,373-1\,677 & $\lambda^2$-fit & 5+1/2\tablefootmark{(c)} & (43) & \\
	\end{tabular}
	\tablefoot{
		\tablefoottext{a}{Three different frequency ranges, 4\,510-8\,490 MHz, 4\,510-8\,300 MHz, 1\,340-4\,960 MHz, were used.}
		\tablefoottext{b}{Compilation of several previously published data sets.}
		\tablefoottext{c}{Data points that seem to be duplications of the same observations, appearing in two different catalogs, are used only once and denoted as half data points for both catalogs, so that the sum of the data points is the total number of data points used.}
		\tablefoottext{d}{Additional data from (31) used in the fit.}
		\tablefoottext{e}{Compilation of several earlier data sets, including an unpublished one for which no details are provided.}
		\tablefoottext{f}{Compilation of several earlier data sets and the one described in (22).}
		\tablefoottext{g}{Three different frequency ranges, 1\,360-1\,690 MHz, 1\,373-4\,898 MHz, 1\,373-4\,990 MHz, were used.}
		\tablefoottext{h}{The frequency range is 4\,736-8\,564 MHz for the ATCA observations and 4\,605-8\,655 MHz for the VLA observations.}
		\tablefoottext{i}{The frequency range is shifted to lower frequencies by up to 40 MHz for some sources.}
		\tablefoottext{j}{RMs calculated from previously published and unpublished data, as well as new measurements with various instruments.}
		\tablefoottext{k}{RMs calculated from previously published polarization data.}
		\tablefoottext{l}{$\lambda^2$-fits were also produced and found to agree with the synthesis results.}
	}
	\tablebib{(1)~\citet{bonafede-2010}; (2)~\citet{broten-1988}; (3)~\citet{brown-2003a}; (4)~\citet{taylor-2003}; (5)~\citet{brown-2007}; (6)~\citet{haverkorn-2006b}; (7)~\citet{mcclure-griffiths-2005}; (8)~\citet{clarke-2001}; (9)~\citet{clarke-2004}; (10)~\citet{clegg-1992}; (11)~\citet{feain-2009}; (12)~\citet{feain-2011}; (13)~\citet{gaensler-2001}; (14)~Hammond (private communication); (15)~\citet{heald-2009}; (16)~\citet{braun-2007}; (17)~\citet{hennessy-1989}; (18)~\citet{johnston-hollitt-2011}; (19)~\citet{johnston-hollitt-2003}; (20)~\citet{johnston-hollitt-2004b}; (21)~\citet{kato-1987}; (22)~\citet{kim-1991}; (23)~\citet{klein-2003}; (24)~\citet{gregorini-1998}; (25)~\citet{vigotti-1999}; (26)~\citet{lawler-1982}; (27)~\citet{dennison-1979}; (28)~\citet{mao-2010}; (29)~\citet{mao-2011}; (30)~\citet{gaensler-2005}; (31)~\citet{mao-2008}; (32)~\citet{minter-1996}; (33)~\citet{oren-1995}; (34)~O'Sullivan (private communication); (35)~\citet{roy-2005}; (36)~\citet{rudnick-1983}; (37)~\citet{schnitzeler-2011}; (38)~\citet{simard-normandin-1981}; (39)~\citet{tabara-1980}; (40)~\citet{taylor-2009}; (41)~\citet{condon-1998}; (42)~\citet{vaneck-2011}; (43)~\citet{wrobel-1993}.}
\end{table}
\end{landscape}

\end{document}